\documentclass[11pt]{article}

\usepackage[final]{acl}

\usepackage{times}
\usepackage{latexsym}

\usepackage[T1]{fontenc}

\usepackage[utf8]{inputenc}

\usepackage{microtype}

\usepackage{inconsolata}

\usepackage{graphicx}

\usepackage{multirow}

\usepackage{booktabs}

\usepackage{tabularx}

\usepackage{ragged2e}

\usepackage{amssymb}   %

\usepackage{graphicx}  %

\usepackage{xcolor}

\usepackage{tcolorbox}

\tcbuselibrary{breakable,listings}

\newtcblisting{promptblock}{
  breakable,
  listing only,
  colback=gray!5,
  colframe=gray!45,
  boxrule=0.4pt,
  arc=1.5pt,
  left=4pt,
  right=4pt,
  top=3pt,
  bottom=3pt,
  boxsep=0pt,
  enhanced,
  listing options={
    basicstyle=\ttfamily\scriptsize,
    breaklines=true,
    columns=fullflexible,
    keepspaces=true,
    showstringspaces=false
  }
}

\usepackage{enumitem}
\usepackage{xspace}
\usepackage{tcolorbox}
\tcbuselibrary{skins}
\usepackage{booktabs}
\usepackage{multirow}
\usepackage{makecell}

\def\tool{\textbf{TeleGen}\xspace}

\title{TeleGen: Improving LLM-Based Web Application Generation via Runtime Telemetry}

\author{
Yujia Luo$^{1}$ \quad
Haonan Zhang$^{2}$ \quad
Jiasi Shen$^{3}$ \quad
Zishuo Ding$^{1,\dagger}$ \quad
Weiyi Shang$^{2}$ \\[-0.1em]
{
$^{1}$The Hong Kong University of Science and Technology (Guangzhou)}\\
$^{2}$University of Waterloo\\
$^{3}$The Hong Kong University of Science and Technology\\
\texttt{yluo246@connect.hkust-gz.edu.cn} \quad
\texttt{haonan.zhang@uwaterloo.ca} \quad
\texttt{sjs@cse.ust.hk} \\
{\texttt{zishuoding@hkust-gz.edu.cn} \quad
\texttt{wshang@uwaterloo.ca} \qquad
}
}

\begin{document}
\maketitle
\begingroup
\renewcommand{\thefootnote}{\fnsymbol{footnote}}
\footnotetext[2]{Corresponding author.}
\endgroup

\begin{abstract}

Large language models can generate runnable web applications from natural-language requirements, but many generated applications still fail interactive tasks. Existing generate-execute-repair pipelines execute the generated application and use task outcomes or error messages to guide code revision. However, this feedback often misses the runtime behavior between a browser action and the final task outcome, making interaction-level failures difficult to diagnose. Therefore, we propose \tool, an observability-enhanced framework for LLM-based web application generation. \tool instruments generated applications, collects runtime telemetry during task execution, and compresses raw telemetry logs into concise briefs for repair. We evaluate \tool on WebGen-Bench and Web-Bench. On WebGen-Bench, \tool improves task success from 67.7\% with repair without telemetry to 76.2\%, an increase of 8.5 percentage points. On Web-Bench, it improves cumulative Pass@2 from 21.7\% to 29.8\%. Ablation results show that runtime telemetry provides a useful diagnostic signal, while telemetry briefs make this signal more effective and less costly to use. Further analysis shows that telemetry is especially helpful for failures involving hidden execution paths, such as navigation, form workflows, and frontend-backend coordination.

\end{abstract}

\section{Introduction}

Large language models (LLMs) have made substantial progress in code generation~\citep{HumanEval, mbpp, CodeGenSurvey, Qwen25Coder}, moving from standalone programming problems~\citep{APPS, AlphaCode, LiveCodeBench, EvalPlus} to repository-level tasks~\citep{RepoBench, RepoCoder, SWEbench, OpenHands}. Within web code generation, the task has similarly shifted from producing static webpages to building interactive web applications from requirements expressed in natural language, screenshots, or domain-specific languages~\citep{WebGenAgent, Interaction2Code, MMWebAgent, UI2CodeN, MultiAgentWebGen, EditWithNL}. Static web page generation asks a model to translate a visual design or screenshot into HTML/CSS that reproduces the intended layout and appearance~\citep{Design2Code, UIFromScreenshots}. Interactive web application generation goes further: the model must produce runnable website code from a high-level natural-language requirement, implementing user-facing workflows such as search, navigation, and form submission~\citep{Interaction2Code}. Correctness therefore depends not only on the rendered appearance or individual code fragments, but also on whether user interactions trigger the right frontend logic,  state updates, and backend behavior during execution~\citep{Crawljax}.

To improve the quality of generated code, recent work adopts a generate-execute-repair workflow~\citep{WebGenBench}. After generating an initial application, an execution step exercises its behavior by performing user-oriented tasks, either through browser agents that interact with the application in realistic web environments~\citep{Mind2Web, WebArena, WebVoyager, VisualWebArena} or through end-to-end testing frameworks that programmatically verify behavior against expected outcomes~\cite{RethinkGUITest, WebTestPilot, WebAgentTesters}. This step exposes failures that the source code alone does not reveal, such as a button that does not trigger its handler~\citep{TestJSApp}, or a form submission that does not update the interface~\citep{TestAjaxUI}. When execution reveals a task failure, the repair agent~\citep{SemAgent, NeuroSymbolicAgent, Xia2023LLMAPR} receives the feedback and revises the generated code.

While effective, this feedback is often incomplete for interactive web applications: interaction-based execution can reveal that a generated application fails a task, but it can miss critical information that is not captured through the visible UI layer alone~\citep{UnderstandJSEvents}. 
As a result, the feedback that reaches the repair agent~\citep{Fan2023RepairLLMPrograms, RepairAgent} is typically limited to task-level pass/fail outcomes or surface-level error messages. For example, after a browser agent clicks a button, the available feedback may not show if the click reached the intended event handler, sent a request, received the expected backend response, or the application state changed but was not rendered~\citep{StudyJSBugs}. From the outside, these failures may look similar as the expected result was not reached. For repair, however, they point to different defects and require different code changes~\citep{JSFaultLocalization}.

This gap motivates our central question: can making generated applications more observable during execution provide better runtime evidence for repair? To answer it, we add an observability step before repair. Drawing on the use of observability to diagnose complex software systems~\citep{Li2022Observability, BenifitCostOfLog}, we instrument the generated application with lightweight logging mechanisms and record its runtime behavior while the browser agent executes each task.  The resulting runtime telemetry~\citep{LogAndTelemetry} captures interaction events, runtime errors, state changes, and other task-relevant execution signals. Rather than passing the raw telemetry logs directly to the repair agent, we employ a separate agent to compress them into a telemetry brief. Raw logs often contain duplicated messages~\citep{Dlfinder, DuplicatedLogStudy}, framework noise, and other low-level details that can obscure the evidence relevant to a failed task~\citep{LogAssist}. They are also lengthy, increasing the context size and cost of each repair attempt~\citep{LostInTheMiddle}. The telemetry brief is designed to preserve concise diagnostic evidence while avoiding the noise and cost of long execution logs.

We evaluate our approach on two web code-generation benchmarks, WebGen-Bench~\citep{WebGenBench} and Web-Bench~\cite{WebBench}, covering both full web application generation and sequential web-development tasks. Across both benchmarks, adding runtime observability to the generation pipeline improves the quality of the final generated applications compared with standard generation and repair without telemetry. On WebGen-Bench, task success improves from 61.7\% for the initially generated applications to 67.7\% after repair without telemetry, and further to 76.2\% when repair is guided by telemetry-derived evidence. At the task level, telemetry-guided generation fixes 106 tasks that the no-telemetry repair baseline fails, while losing 51 tasks solved by the baseline, giving a net gain of 55 tasks. On Web-Bench, our framework improves Pass@2 from 21.7\% to 29.8\%. Our further analysis shows that runtime evidence is especially useful for failures involving navigation, form and input workflows, authentication, search and filtering, and report or export behavior.

Our contributions can be summarized as follows:
\begin{itemize}
    \item We identify an observability gap in interactive web applications pipelines: existing feedback often misses the runtime behavior behind task failures.
    \item We propose and evaluate an observability-enhanced framework that instruments generated applications, collects runtime logs, and compresses them into telemetry briefs for repair, improving final application quality on WebGen-Bench and Web-Bench.
    \item We show that telemetry is especially useful for failures involving hidden execution paths such as interaction handling, state updates, navigation, network behavior, and frontend-backend coordination.
\end{itemize}

\paragraph{Availability.}
The code, prompts, and experimental results are publicly available on
\href{https://github.com/commoluo/TeleGen}{GitHub}.

\begin{figure*}[!t]
    \centering
    \includegraphics[width=\textwidth]{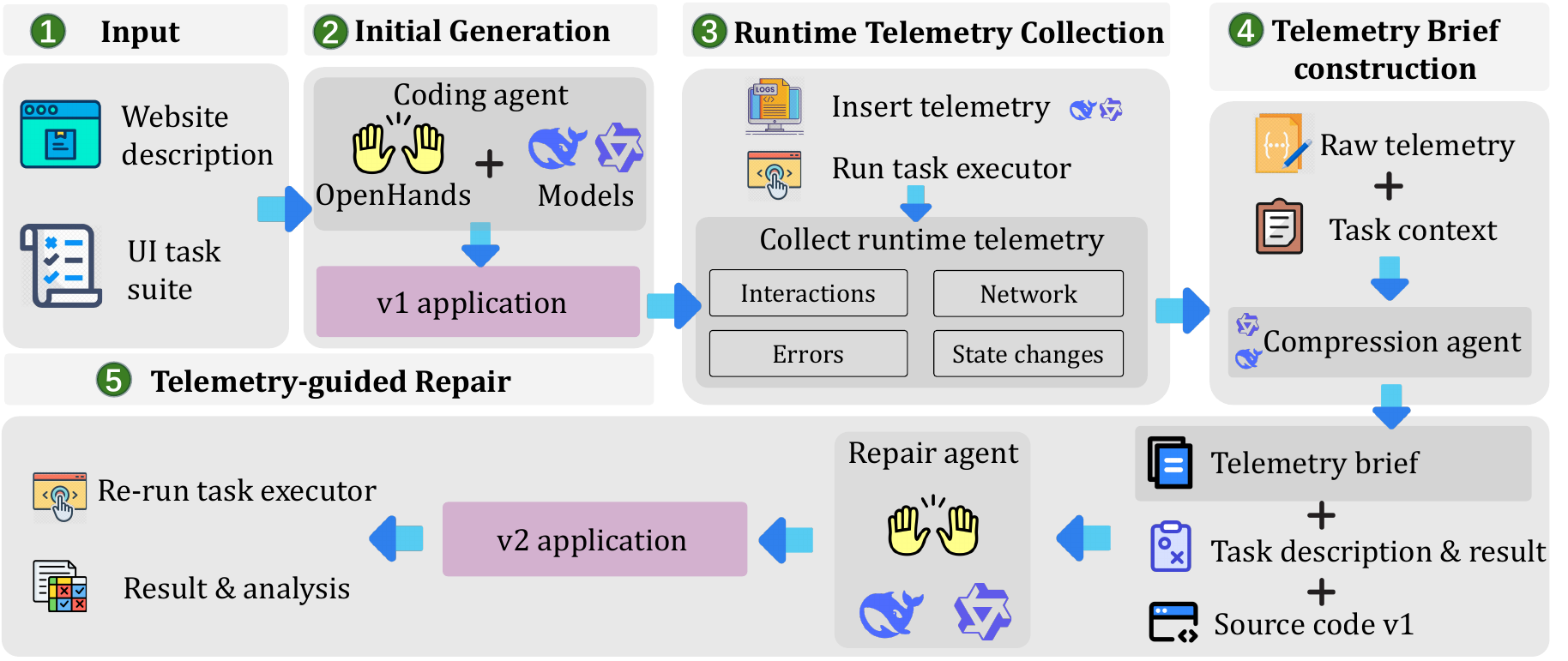}
    \caption{Overview of \tool. Given a website description and UI task suite, a coding agent first generates an initial web application, denoted as the \textit{v1 application}. \tool then instruments a separate copy of the v1 application and re-runs the task executor to collect runtime telemetry, including actions, network events, errors, and state changes. A compression agent combines the raw traces with the task context and produces a compact telemetry brief. Finally, the repair agent edits the clean v1 source code using the telemetry brief and task feedback, producing a repaired \textit{v2 application}, which is re-run by the task executor for final result analysis.}
\label{fig:overview}
\end{figure*}

\section{Related Work}
\label{sec:related_work}

\paragraph{LLM-based web application generation.}
Recent work has studied generating webpages and website codebases from screenshots, design inputs, interaction traces, and natural language requirements~\citep{Design2Code, Web2Code, Interaction2Code, WebGenBench, WebGenAgent, MMWebAgent}. These tasks move beyond standalone code generation because the output must render a plausible interface and support user behavior through routing, events, state updates, and backend logic. \citet{Design2Code} and \citet{Web2Code} study front end code generation from visual inputs, while \citet{Interaction2Code} shows that interactive behavior remains difficult to generate reliably. \citet{WebGenBench} is closest to our study because it evaluates whether generated web applications can complete functional UI tasks. We build on this setting by studying the next step, repairing the generated application after browser evaluation exposes failed workflows.

\paragraph{Web agents and browser-based evaluation.}
Web agents have been widely used to test whether language models can complete user-oriented tasks through browser interaction~\citep{WebGUM, Mind2Web, WebArena, WebVoyager, VisualWebArena, WorkArena, BrowserGym, Mind2Web2, WebCoT}. These benchmarks evaluate behavior through execution rather than static inspection. In web application generation, browser agents can also act as testers by interacting with a generated application and checking whether the expected result is reached~\citep{WebGenBench}. This task completion signal is useful, but it is often too coarse for repair. It can show that a workflow failed, but not whether the cause was an inactive event handler, a failed request, a missing route transition, or a state update that was not reflected in the interface. We study whether runtime evidence collected during browser execution can provide this missing diagnostic signal.

\paragraph{LLM-based program repair.}
Large language models have been studied for automated program repair, including repairing incorrect programs produced by LLMs, generating patches for standard repair benchmarks, and resolving repository-level software issues~\citep{Fan2023RepairLLMPrograms, Xia2023LLMAPR, SWEbench}. Recent repair systems further treat the model as an agent that can gather context, reason about defects, use tools, and validate candidate fixes~\citep{RepairAgent, SemAgent}. Generated web applications pose a related but different repair problem. The target is an interactive system verified through browser workflows, and the failure cause may only become visible during execution. Our work complements prior repair research by making the generated application observable during evaluation and by studying whether telemetry evidence improves repair beyond source code, task descriptions, and final outcomes.

\section{Approach}
\label{sec:approach}

To study how runtime telemetry can improve LLM-generated web applications, we propose \tool, an observability-enhanced framework for web application generation. Figure~\ref{fig:overview} illustrates the pipeline. Given a website description, a coding agent first generates an initial application, which we call the \textit{v1 application}. A task executor then runs the corresponding task suite on this application and produces task feedback, including pass/fail outcomes. Then, \tool adds an observability step, where it creates an instrumented copy of the \textit{v1 application} and runs the same tasks on this copy. During execution, the instrumented application records runtime telemetry. These raw telemetry logs are then compressed into a telemetry brief. The repair agent receives the clean \textit{v1} source code, task feedback, and, in the \tool setting, the telemetry brief. It then edits the clean source code to produce a revised \textit{v2 application}. Finally, the task executor re-runs the same task suite on the \textit{v2 application} to measure whether the final generated application better satisfies the intended workflows.

\subsection{Initial Generation and Task Execution}
\label{sec:initial-generation-evaluation}

\tool augments existing web code-generation pipelines rather than replacing their original generation or execution protocols. For each benchmark, we therefore follow the setup used in the corresponding original work (cf. Section~\ref{sec:experiments}). For example, in WebGen-Bench~\citep{WebGenBench}, we use the OpenHands-based~\citep{OpenHands} generation setting to produce applications from natural-language website specifications, then execute the generated applications with the benchmark’s WebVoyager~\citep{WebVoyager} UI-task protocol. Web-Bench~\citep{WebBench} is handled analogously by following its original Web Agent and Playwright-based~\citep{Playwright} testing workflow.

In these benchmarks, the original execution step already provides simple feedback. This feedback may come from task outcomes, build failures~\citep{BuildFailures}, assertion mismatches~\citep{OracleMismatch}, or test errors from automation frameworks~\citep{RethinkGUITest}. \tool keeps this feedback information unchanged at this stage. This design lets us study whether observability improves existing generate-execute-repair pipelines without changing the generator, executor, or benchmark protocol.

\subsection{Runtime Telemetry Collection}
\label{sec:making-observable}

To collect runtime telemetry, \tool instruments a separate copy of the v1 application while preserving the clean v1 source for later repair. This separation keeps telemetry collection independent from the repair step and avoids confounding the repair results with possible side effects introduced during instrumentation. 

Specifically, the instrumentation is performed by an LLM-based coding agent. Given the application source, website specification, and task suite, the agent inserts lightweight logging statements at code locations that it judges useful for diagnosing task failures. This design makes the logging repair-oriented: since the telemetry is ultimately consumed by a coding agent during repair, we ask a coding agent to decide what execution signals are likely to be useful. The prompt explicitly instructs the agent to add telemetry only, without changing application behavior.

After instrumentation, the task executor runs the same tasks on the instrumented application. During this run, the inserted logging statements record runtime telemetry and capture how browser actions propagate through the application. This makes intermediate application behavior visible, instead of leaving the repair agent with only the original task feedback.

\subsection{Telemetry Brief Construction}
\label{sec:telemetry-compression}

The raw telemetry logs collected during task execution are not passed directly to the repair agent. Although they contain complete runtime details, they also include duplicated messages or other types of noise. Long logs further increase context cost and can make the useful evidence harder to identify~\citep{LostInTheMiddle}.

\tool therefore uses a separate compression agent to convert raw telemetry logs into a telemetry brief. The agent receives the task context together with the collected logs, then preserves execution behavior that is likely to explain the failure while removing repeated or irrelevant entries. The brief is not intended to be a complete execution transcript. Instead, it gives the repair agent a compact, repair-oriented view of what happened during task execution. For example, it may show that a request succeeded but no visible update followed, or that data was loaded but no selection event occurred. Such evidence helps distinguish failures that appear similar from the final task outcome alone.

\subsection{Telemetry-Guided Repair}
\label{sec:observable-repair}

After the telemetry brief is constructed, the repair agent receives the clean v1 source code together with the task description, expected result, pass/fail outcome and telemetry brief. It edits the clean v1 application rather than the instrumented copy, producing the \textit{v2 application}. This separation ensures that instrumentation is used only for observation, while functional changes are made during repair. Guided by the telemetry brief, the repair agent can localize the likely cause of a failure before modifying the source code. 

The v2 application is then executed with the same task suite and evaluation protocol as the original pipeline. Comparing this setting with standard repair measures whether runtime telemetry improves the final generated application.

\section{Experimental Settings}
\label{sec:experiments}
We evaluate whether runtime telemetry improves LLM-based web application generation. This section describes the benchmarks, agents, models, and metrics used in our experiments. The main comparisons follow the original baselines of each benchmark (cf. Section~\ref{sec:initial-generation-evaluation}), while the component-level ablation is described in Section~\ref{sec:results-rq2}.

\subsection{Benchmarks and Task Setting}
\label{sec:benchmarks-task-setting}

We evaluate \tool on two web code-generation benchmarks with different task formats. WebGen-Bench evaluates full web application generation from natural-language requirements. The input is a website specification describing the intended application and its user-facing functionality, while the expected output is a runnable multi-file web application. The generated application is tested with UI tasks, where each task contains a user goal and an expected final result. WebGen-Bench contains 101 projects and 647 UI tasks, which are grouped into Content Presentation (174 tasks), User Interaction (313 tasks), and Data Management (160 tasks).

Web-Bench evaluates sequential web development tasks. Each project contains an initial codebase and a sequence of dependent tasks. At each step, the input is the current project state together with the next task description, and the expected output is a set of code edits that implement the requested feature. The result is checked with end-to-end tests. Web-Bench contains 50 projects, each consisting of 20 tasks. We use it to examine whether runtime telemetry remains useful when the benchmark already includes a second attempt with test-error feedback. The benchmark also provides three difficulty groups: Easy (277 tasks), Moderate (236 tasks), and Challenging (487 tasks).

\subsection{Agents and Models}
\label{sec:agents-models}

For each benchmark, we follow its original code-agent framework and execution setup. On WebGen-Bench, we use the OpenHands-based generation setting and execute UI tasks with WebVoyager. WebVoyager interacts with the generated application through the browser and judges whether each expected result is reached. On Web-Bench, we use its own Web-Agent framework, which generates file edits for each development task and relies on Playwright-based end-to-end tests to check results.

Our main experiments use DeepSeek-V4-Flash as the text/code model for generation, telemetry insertion, telemetry-brief construction and repair. Additional results with DeepSeek-V4-Pro, Qwen3.5-Plus and Gemini-3-Flash-Preview are reported in Appendix A. For WebGen-Bench, all WebVoyager model-driven steps use Qwen3.5-Plus.

\subsection{Metrics and Analysis}
\label{sec:metrics}

For WebGen-Bench, our main metric is task success rate, defined as the fraction of UI tasks judged successful by the browser-agent evaluator. We report overall task success and category-level success. We also report token usage to measure the context cost introduced by runtime telemetry and telemetry-brief construction.

For Web-Bench, we follow the original benchmark and report Pass@1 together with cumulative Pass@2. Pass@1 measures the fraction of tasks that pass all end-to-end tests on the first attempt. Pass@2 measures the fraction of tasks passed after the retry step, where failed tasks receive error context and the agent is given a second attempt. Because \tool only changes the second attempt in Web-Bench, Pass@1 remains unchanged between the original and telemetry-augmented settings.

For paired task-level comparisons, we use two-sided exact McNemar tests on the discordant outcomes between settings.

We further analyze tasks solved by \tool but not by the no-telemetry baseline to identify where runtime telemetry is most useful. These cases are grouped by failure type to study which kinds of web-application failures benefit most from runtime telemetry.

%

%

%

%

%
%

%
%
%
%

%
%
%

%
%
%
%

%
%

%

%

%

%

%
%
%
%
%
%
%
%
%
%
%
%
%
%
%
%
%
%
%
%
%
%
%
%
%
%
%
%
%
%
%
%
%
%
%
%
%
%
%
%
%
%
%
%
%
%
%

%

%
%
%
%
%
%
%
%
%
%
%
%
%
%
%
%
%
%
%
%
%
%
%
%
%
%
%
%

\begin{table*}[t]
\centering
\small
\renewcommand{\arraystretch}{1.10}
\begin{tabular}{c c c c c @{\hspace{1.4em}} c c c}
\toprule
\multirowcell{2}{\textbf{Benchmark}}
& \multirowcell{2}{\textbf{Setting}}
& \multirowcell{2}{\textbf{Passed}\\\textbf{/ Total}}
& \multirowcell{2}{\textbf{Overall}}
& \multirowcell{2}{\textbf{$\Delta$}}
& \multicolumn{3}{c}{\textbf{Categories}} \\
\cmidrule(lr){6-8}
& & & & & \textbf{Content} & \textbf{User} & \textbf{Data} \\
\midrule
\multirowcell{3}{WebGen-Bench}
& Initial generation & 399 / 647 & 61.7 & -- & 71.3 & 55.0 & 64.4 \\
& No-Telemetry repair & 438 / 647 & 67.7 & +6.0 & 78.7 & 60.4 & 70.0 \\
& \textbf{\tool} & 493 / 647 & \textbf{76.2} & \textbf{+14.5} & \textbf{88.5} & \textbf{68.1} & \textbf{78.8} \\
\midrule
& & & & & \textbf{Easy} & \textbf{Moderate} & \textbf{Challenging} \\
\midrule
\multirowcell{3}{WebBench}
& Pass@1 & 142 / 1000 & 14.2 & -- & 44.0 & 7.2 & 0.6 \\
& Original Pass@2 & 217 / 1000 & 21.7 & +7.5 & 59.6 & 16.9 & 2.5 \\
& \textbf{\tool{} Pass@2} & 298 / 1000 & \textbf{29.8} & \textbf{+15.6} & \textbf{72.9} & \textbf{32.2} & \textbf{4.1} \\
\bottomrule
\end{tabular}
\caption{For WebGen-Bench, Overall, Content, User and Data report task-level success rates for the full benchmark and its three project categories. For Web-Bench, Pass\@1 is the first-attempt success rate, while Original Pass\@2 and \tool{} Pass\@2 report cumulative success after the original and telemetry-augmented second attempts. Easy, Moderate and Challenging report category-level cumulative success rate. For WebGen-Bench, $\Delta$ denotes the overall improvement of \tool{} over Initial generation; for Web-Bench, $\Delta$ denotes the overall improvement over Pass\@1. All values are percentages except passed-task counts.
}
\label{tab:rq1-main-results}
\end{table*}

\section{Results and Analysis}
\label{sec:results}

\subsection{Effectiveness of \tool}
\label{sec:results-rq1}
Overall, \tool improves the final generated applications on both benchmarks, showing that runtime telemetry is useful across both full web application generation and sequential web-development tasks.
Table~\ref{tab:rq1-main-results} reports the main results under the DeepSeek-V4-Flash configuration. On WebGen-Bench, the initially generated applications pass 399 out of 647 tasks, corresponding to 61.7\% task success. Standard one-round repair without telemetry increases this to 438 tasks, or 67.7\%. With runtime telemetry, \tool{} further improves task success to 493 tasks, or 76.2\%, yielding an 8.5 percentage-point gain over the matched No-Telemetry repair baseline. The paired improvement over No-Telemetry repair is statistically significant under a two-sided exact McNemar test ($p=1.35\times10^{-5}$).

The improvement is consistent across all three WebGen-Bench project categories. Compared with No-Telemetry repair, \tool{} improves success by 9.8 points on Content Presentation (78.7\% to 88.5\%), 7.7 points on User Interaction (60.4\% to 68.1\%), and 8.8 points on Data Management (70.0\% to 78.8\%). These gains suggest that runtime telemetry provides useful repair evidence across different kinds of generated web applications rather than only one project type.

Because WebGen-Bench relies on WebVoyager for outcome judgment, we additionally conduct an outcome-blinded human audit of 60 sampled tasks. Human judgments agree with WebVoyager on 54 of 60 cases (90.0\%), with Cohen's $\kappa=0.80$. The full audit protocol is reported in Appendix~\ref{sec:webvoyager-audit}.

Web-Bench provides a complementary setting based on sequential web-development tasks. The first-attempt Pass@1 is 14.2\%, with 142 out of 1000 tasks passing before retry. The original Web-Bench retry pipeline improves cumulative Pass@2 to 21.7\%. When the second attempt is augmented with runtime telemetry, \tool further improves cumulative Pass@2 to 29.8\%. This corresponds to 81 additional passed tasks over the original retry pipeline and a 15.6\% gain over the first attempt. The paired improvement over Original Pass@2 is also statistically significant under a two-sided exact McNemar test ($p=4.49\times10^{-10}$).

The category-level results show that the improvement is not limited to easier tasks. On Easy tasks, \tool improves cumulative Pass@2 from 59.6\% to 72.9\%, suggesting that telemetry helps the agent recover failures even when the underlying task is relatively accessible. On Moderate tasks, the gain is larger in absolute terms, rising from 16.9\% to 32.2\%, which indicates that runtime telemetry is particularly useful when the model needs more diagnostic information to decide how to repair the code. Challenging tasks remain difficult overall, but \tool still improves performance from 2.5\% to 4.1\%. These results suggest that telemetry strengthens the original Web-Bench retry pipeline by providing additional execution context beyond standard test-error feedback, with the clearest benefit on tasks where failures are recoverable but not obvious from error messages alone.

\subsection{Ablation Studies}
\label{sec:results-rq2}

We conduct a step-by-step ablation of \tool to analyze how runtime telemetry and telemetry-brief construction contribute to performance. We perform the ablation on WebGen-Bench, where all variants start from the same generated v1 applications and use the same one-round repair protocol. Table 2 reports the results.

The first variant is no-telemetry repair, where the repair agent receives the standard repair inputs, including the clean v1 source code, task description, expected result and pass/fail outcome. The second variant is no-brief repair, where the repair agent additionally receives raw telemetry logs collected from the instrumented v1 application. The final variant is \tool, where the raw telemetry logs are compressed into telemetry briefs before repair.

Compared with no-telemetry repair, adding raw telemetry logs increases the number of passed tasks from 438 to 484, improving task success from 67.70\% to 74.81\%. This 7.11-point gain shows that the runtime signals collected by \tool contain diagnostic information that is not available from the standard repair inputs alone. In other words, the repair agent benefits from observing what happened during execution, not only from knowing the final task outcome.

Compressing raw logs into telemetry briefs further improves success to 493 passed tasks, or 76.20\%. The gain over raw logs is smaller than the gain from adding telemetry itself, but it is still meaningful because both settings use the same underlying runtime information. The difference is how that information is presented to the repair agent. This suggests that telemetry is most useful when it is organized into a concise, task-centered form rather than passed as long raw logs.

Table~\ref{tab:rq2-token-cost} further shows that telemetry-brief construction improves the effectiveness-cost tradeoff. Compared with no-brief repair, \tool reduces average total token cost from 1481.7K to 1458.1K tokens per project while improving task success rate. Across the full run, this saves 2.39M tokens, or 1.6\%. Overall, the ablation shows that runtime telemetry is the primary source of improvement, while telemetry briefs help the repair agent use this information more effectively and with lower context cost.

Beyond token usage, the additional monetary overhead of runtime telemetry is small under the model pricing used in our experiments. TeleGen costs approximately \$0.41 per project, compared with \$0.28 for No-Telemetry repair, while improving task success from 67.7\% to 76.2\%. Detailed cost accounting and runtime measurements are reported in Appendix~\ref{sec:cost-analysis}.

\begin{table}[t]
\centering
\small
\renewcommand{\arraystretch}{1.12}
\setlength{\tabcolsep}{5pt}
\begin{tabular}{lrrr}
\toprule
\textbf{Setting} & \textbf{Passed} & \textbf{Success} & \textbf{$\Delta$} \\
\midrule
No-telemetry & 438 & 67.70\% & -- \\
No-brief & 484 & 74.81\% & +7.11 pp \\
\tool & 493 & 76.20\% & +8.50 pp \\
\bottomrule
\end{tabular}
\caption{Repair effectiveness under DeepSeek-V4-Flash. $\Delta$ is measured against no-telemetry repair.}
\label{tab:rq2-repair-effectiveness}
\end{table}

\begin{table}[t]
\centering
\small
\setlength{\tabcolsep}{5pt}
\begin{tabular}{lrrr}
\toprule
\textbf{Setting} & \textbf{Input} & \textbf{Output} & \textbf{Total} \\
\midrule
No-telemetry & 971.8K & 19.0K & 990.8K \\
No-brief & 1462.1K & 19.6K & 1481.7K \\
\tool & 1439.6K & 18.4K & 1458.1K \\
\bottomrule
\end{tabular}
\caption{Average token cost per project under DeepSeek-V4-Flash. Values are reported in thousands of tokens.}
\label{tab:rq2-token-cost}
\end{table}

\begin{figure}[t]
    \centering
    \includegraphics[width=\linewidth]{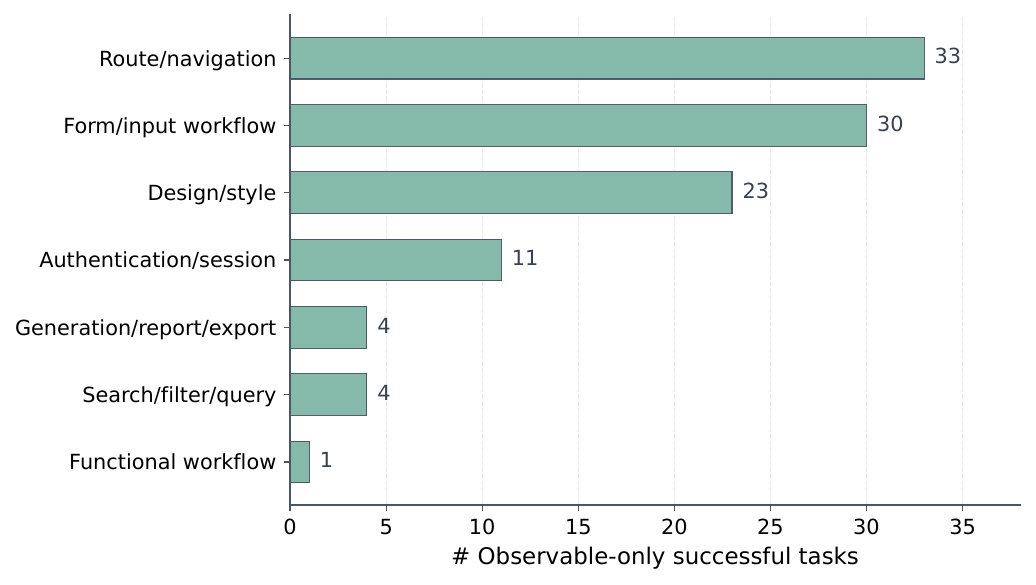}
    \caption{Failure types among tasks solved only by Telemetry-Guided Repair under DeepSeek-V4-Flash. These tasks fail under No-telemetry repair but succeed when telemetry brief is provided.}
    \label{fig:rq3-failure-types}
\end{figure}
\subsection{Failures  Helped by Runtime Telemetry}
\label{sec:results-rq3}

\begin{figure*}[t]
    \centering
    \includegraphics[width=\textwidth]{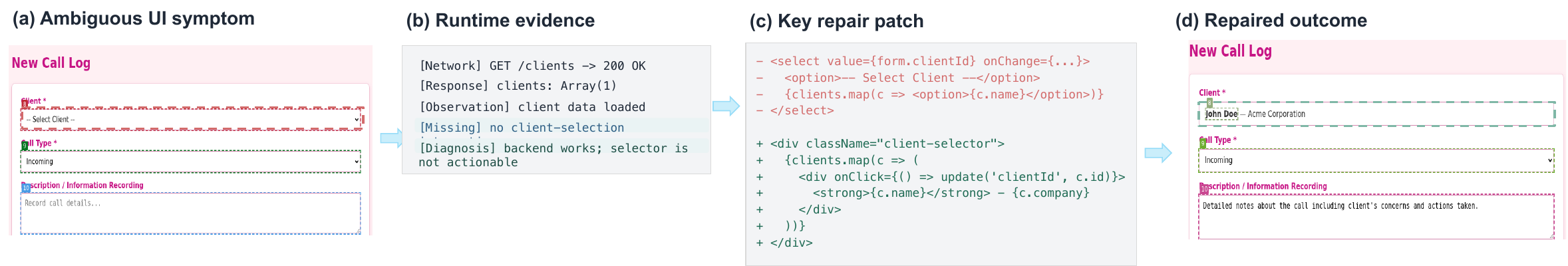}
    \caption{Representative form/input workflow repair enabled by runtime evidence. Although the form appears valid, runtime telemetry shows that client data is loaded but no client-selection interaction occurs. Telemetry-Guided Repair replaces the hard-to-operate dropdown with explicit clickable client cards, allowing the browser agent to complete the workflow.}
    \label{fig:rq3-case-study}
\end{figure*}
To understand where runtime telemetry helps, we analyze the 106 tasks solved by \tool but not by no-telemetry repair. These tasks provide a direct view of the added value of telemetry, since both settings start from the same generated applications and use the same repair protocol, while only \tool receives telemetry-derived execution information.

Figure~\ref{fig:rq3-failure-types} shows the distribution of these \tool-only successes. Route/navigation failures form the largest group, with 33 tasks. These failures are difficult to diagnose from final outcomes because the repair agent may only know that the expected page or view did not appear. Runtime telemetry can reveal whether the click reached the intended handler, whether the route update was attempted, or whether the application remained on the same page after interaction. This helps the repair agent distinguish missing navigation links from broken routing logic or state-dependent redirects.

Form/input workflow failures are the second largest group, with 30 tasks. These failures often involve several intermediate steps between user input and visible confirmation. A form may render correctly, but the submitted values may not be captured, validation may reject the input, or the submit handler may fail to update the interface. Runtime telemetry exposes where the workflow stops, making it easier to decide whether the fix should target input binding, validation rules, handler logic or rendering after submission.

Design/style failures account for 23 tasks. Although these failures appear visual, many are tied to dynamic UI behavior rather than static appearance alone. For example, a required element may be missing only after filtering, navigation or state changes. Telemetry helps determine whether the application produced the expected state internally but failed to render it, or whether the relevant state was never reached.

Authentication/session failures account for 11 tasks. These failures often depend on coordination between frontend state and backend responses. Runtime telemetry can reveal whether login actions sent the expected request, whether session state changed after the response, or whether protected content failed to render after authentication. This is especially useful because the final UI may simply show that the user remains unauthenticated, without explaining where the authentication flow broke.

The remaining \tool-only successes involve generation/report/export, search/filter/query and general functional workflows. These categories are smaller, but they follow the same pattern: telemetry helps when the failure occurs between the browser action and the visible result. For export tasks, it can show whether the generation action ran but no artifact became available. For search and filtering, it can reveal whether the query state changed, the filtering logic executed, or the result list failed to update.

Overall, the failure-type analysis shows that runtime telemetry is most useful when task failure hides multiple possible causes. By exposing intermediate execution behavior, telemetry helps the repair agent distinguish failures that look similar from the outside but require different code changes.

%

\paragraph{Representative case study.}
Figure~\ref{fig:rq3-case-study} shows a representative form/input workflow failure from the CRM call logging application. The task requires the browser agent to create a new call log by selecting a client and filling in the required fields. The visible UI alone is ambiguous: the form is rendered, the client selector appears available, and the static code contains a standard \texttt{<select>} control with an \texttt{onChange} handler populated from the client API. From task-level feedback alone, the repair agent only knows that the task is not completed; it cannot determine whether the failure comes from data loading, navigation, form submission, state update, or the client-selection control.

Runtime evidence exposes the missing interaction step. The trace shows that the agent reaches the call-log creation page and that the client-loading request succeeds, returning valid client data. However, the execution timeline then contains no follow-up client-selection interaction. This ``silence'' after successful data loading localizes the failure to the frontend affordance: the backend is functional, but the native dropdown is not sufficiently actionable for the browser agent. Guided by this evidence, \tool{} replaces the \texttt{<select>} dropdown with explicit clickable client cards and direct \texttt{onClick} handlers. After this change, the browser agent can select a client and complete the call-log creation workflow.

Overall, the results show that observability is most useful when the cause of failure lies between the user action and the final UI state. Runtime evidence helps bridge this gap by exposing hidden interaction paths, state transitions, handler execution, and frontend--backend behavior that task-level feedback alone does not reveal.

\section{Conclusion}
\label{sec:conclusion}

We presented \tool{} for LLM-generated web applications. Instead of relying only on task-level pass/fail feedback, our approach instruments generated applications, collects runtime evidence during browser-based execution, compresses the resulting traces into a repair-oriented telemetry brief, and uses this evidence to guide code repair. Experiments on WebGen-Bench show that \tool{} improves task success over both initial generation and no-telemetry repair under the main DeepSeek-V4-Flash configuration. Category-level results further show gains across content presentation, user interaction, and data management projects. Our ablation demonstrates that runtime logs provide useful repair signal, while telemetry-brief compression further improves success and reduces token usage compared with using logs directly. Finally, failure-type analysis and a representative case study show that runtime evidence is especially useful for failures involving hidden interaction paths, form workflows, state transitions, and frontend--backend coordination. Overall, our results suggest that making generated web applications observable is a practical way to turn opaque browser-level failures into actionable evidence for LLM-based code repair.

\section{Limitations}
\label{sec:limitations}

While our results suggest that observability provides useful runtime evidence for repairing generated web applications, our current implementation has several limitations. 

First, the observability layer relies on LLM-based telemetry instrumentation. Although we constrain the instrumentation model to output patches and apply these patches locally, the process can still introduce variance. The model may miss important execution points, insert noisy or redundant logs, or modify code locations that are already correct. In rare cases, instrumentation may affect application behavior rather than only exposing it. Our telemetry sanitization and compression step reduces the impact of noisy logs, but it does not fully eliminate the risk of incomplete or imperfect instrumentation.

Second, our current instrumentation strategy is intentionally conservative. The model inserts telemetry based on the source code and task context in order to expose as much potentially relevant runtime behavior as possible. This broad insertion strategy can increase runtime overhead, produce long execution logs, and consume additional storage and context budget. While the telemetry brief reduces the amount of evidence eventually provided to the repair agent, it does not remove the cost of collecting and storing the raw traces. More selective or adaptive instrumentation strategies could reduce this overhead in future work.

Despite these limitations, the results show that making generated web applications observable provides useful repair evidence across a diverse set of browser-based tasks. Future work could explore more deterministic instrumentation, adaptive logging policies, and lightweight runtime tracing mechanisms to further improve the reliability and efficiency of observable repair.

\section*{Acknowledgments}

This work is supported by the Guangdong Science and Technology Department
(Internal Project No. G\_2025\_082).

\bibliography{custom}

\appendix
\appendix

\section{Additional Model Results}
\label{app:model-generalization}

The main paper focuses on DeepSeek-V4-Flash as the primary text/code model configuration.
In this appendix, we report additional model-level results to examine whether the benefit of runtime telemetry is specific to the main configuration.
Table~\ref{tab:app-model-results} includes additional WebGen-Bench results with DeepSeek-V4-Pro and Qwen3.5-Plus as text/code models, as well as an additional Web-Bench validation result under Qwen3.5-Plus.

On WebGen-Bench, \tool achieves the highest success rate for both additional text/code models.
For DeepSeek-V4-Pro, \tool improves overall success from 66.8\% for initial generation and 68.5\% for no-telemetry repair to 73.6\%, increasing the number of passed tasks from 432 and 443 to 476 out of 647 tasks.
For Qwen3.5-Plus, \tool improves overall success from 50.1\% for initial generation and 54.3\% for no-telemetry repair to 60.0\%, increasing the number of passed tasks from 324 and 351 to 388 out of 647 tasks.
For Gemini-3-Flash-Preview, \tool{} improves overall success from 66.3\% under No-Telemetry repair to 69.2\%, corresponding to 19 additional successful tasks and a 2.9 percentage-point gain.
The same trend also holds across the WebGen-Bench project categories: \tool achieves the best result on Content Presentation, User Interaction, and Data Management for both additional text/code models.

The additional Web-Bench result further supports this trend under Qwen3.5-Plus.
Compared with Original Pass@2, \tool{} Pass@2 improves cumulative success from 21.9\% to 24.2\%, increasing the number of passed tasks from 219 to 242 out of 1000 tasks.
The improvement is also consistent across difficulty groups, with Easy increasing from 58.5\% to 61.0\%, Moderate from 19.1\% to 22.0\%, and Challenging from 2.5\% to 4.3\%.

Overall, these additional results suggest that the benefit of runtime telemetry is not limited to the main DeepSeek-V4-Flash configuration.
Across additional text/code models and an auxiliary Web-Bench validation setting, \tool preserves the same overall trend of improving repair outcomes over the corresponding non-telemetry baselines.

\begin{table*}[t]
\centering
\small
\renewcommand{\arraystretch}{1.10}
\setlength{\tabcolsep}{3.4pt}
\begin{tabular}{@{}c l l c r r @{\hspace{0.6em}} r r r@{}}
\toprule
\multirowcell{2}{\textbf{Benchmark}}
& \multirowcell{2}{\textbf{Model}}
& \multirowcell{2}{\textbf{Setting}}
& \multirowcell{2}{\textbf{Passed}\\\textbf{/ Total}}
& \multirowcell{2}{\textbf{Overall}}
& \multirowcell{2}{\textbf{$\Delta$}}
& \multicolumn{3}{c}{\textbf{Categories}} \\
\cmidrule(lr){7-9}
& & & & & & \textbf{Content} & \textbf{User} & \textbf{Data} \\
\midrule

\multirowcell{9}{WebGen-Bench}
& \multirowcell{3}{DeepSeek-\\V4-Pro}
& Initial generation
& 432 / 647
& 66.8 & -- & 85.6 & 55.9 & 67.5 \\
&
& No-telemetry repair
& 443 / 647
& 68.5 & +1.7 & 83.9 & 61.0 & 66.2 \\
&
& \textbf{\tool}
& 476 / 647
& \textbf{73.6} & \textbf{+6.8}
& \textbf{88.5} & \textbf{66.5} & \textbf{71.2} \\

\addlinespace[0.2em]
& \multirowcell{3}{Qwen3.5-Plus}
& Initial generation
& 324 / 647
& 50.1 & -- & 71.8 & 43.1 & 40.0 \\
&
& No-telemetry repair
& 351 / 647
& 54.3 & +4.2 & 75.9 & 47.0 & 45.0 \\
&
& \textbf{\tool}
& 388 / 647
& \textbf{60.0} & \textbf{+9.9}
& \textbf{81.0} & \textbf{54.0} & \textbf{48.8} \\

\addlinespace[0.2em]
& \multirowcell{3}{Gemini-3-\\Flash-Preview}
& Initial generation
& 396 / 647
& 61.2 & -- & 75.9 & 52.4 & 62.5 \\
&
& No-telemetry repair
& 429 / 647
& 66.3 & +5.1 & 78.2 & 58.1 & \textbf{69.4} \\
&
& \textbf{\tool}
& 448 / 647
& \textbf{69.2} & \textbf{+8.0}
& \textbf{82.2} & \textbf{62.3} & 68.8 \\

\midrule
& & & & & & \textbf{Easy} & \textbf{Moderate} & \textbf{Challenging} \\
\midrule

\multirowcell{3}{Web-Bench}
& \multirowcell{3}{Qwen3.5-Plus}
& Pass\@1
& 150 / 1000
& 15.0 & -- & 45.8 & 6.4 & 1.6 \\
&
& Original Pass\@2
& 219 / 1000
& 21.9 & +6.9 & 58.5 & 19.1 & 2.5 \\
&
& \textbf{\tool{} Pass\@2}
& 242 / 1000
& \textbf{24.2} & \textbf{+9.2}
& \textbf{61.0} & \textbf{22.0} & \textbf{4.3} \\

\bottomrule
\end{tabular}

\caption{Additional model results. For WebGen-Bench, Overall, Content, User, and Data report task-level success rates for the full benchmark and three project categories. For Gemini-3-Flash-Preview, all rates use the canonical 647-task denominator, with missing evaluations counted as unsuccessful. For Web-Bench, Pass\@1 is the first-attempt success rate, while Original Pass\@2 and \tool{} Pass\@2 report cumulative success after the original and telemetry-augmented second attempts. Easy, Moderate, and Challenging report category-level cumulative success rate. $\Delta$ denotes the overall improvement over Initial generation for WebGen-Bench and over Pass\@1 for Web-Bench. All values are percentages except passed-task counts.}

\label{tab:app-model-results}
\end{table*}

\section{Additional RQ2 Analysis}
\label{app:rq2-details}

The main text reports the primary RQ2 results, including repair effectiveness and token cost. Here we provide a finer-grained comparison between Observable repair and the two baselines under the main DeepSeek-V4-Flash configuration. We summarize the comparison at both project and task levels.

\subsection{Paired Significance Analysis}

We further assess the task-level differences using two-sided exact
McNemar tests. On WebGen-Bench, \tool{} solves 106 tasks that
No-Telemetry repair fails, while No-Telemetry repair solves 51 tasks
that \tool{} fails. This paired difference is statistically significant
($p=1.35\times10^{-5}$). In contrast, Raw-Telemetry and \tool{} have
56 and 65 exclusive successes, respectively, and their difference is
not statistically significant ($p=0.4672$). On Web-Bench, Original
Pass\@2 and \tool{} Pass\@2 have 45 and 126 exclusive successes,
respectively, yielding a statistically significant difference
($p=4.49\times10^{-10}$).

\subsection{Human Audit of WebVoyager Evaluation}
\label{sec:webvoyager-audit}

Because WebGen-Bench relies on WebVoyager for task-outcome
judgment, we conduct an independent human audit to assess the
reliability of the automatic evaluation. The audit population
contains 493 WebVoyager-judged successes and 154 judged failures
under \tool{}. We use balanced stratified sampling without
replacement, selecting 30 judged successes and 30 judged failures
with random seed 20260713. We select at most one task per project,
resulting in 60 tasks from 60 distinct projects, with no manual
replacement.

One independent human annotator reviews the task description,
expected outcome, execution trajectory, actions, observations,
screenshots, final screenshot, and final application state. The
WebVoyager binary outcome and evaluator response are hidden during
annotation.

\begin{table}[t]
\centering
\footnotesize
\setlength{\tabcolsep}{4.0pt}
\renewcommand{\arraystretch}{1.08}
\begin{tabular}{lrrr}
\toprule
 & \textbf{Human +} & \textbf{Human --} & \textbf{Total} \\
\midrule
WV success & 28 & 2  & 30 \\
WV failure & 4  & 26 & 30 \\
\midrule
Total      & 32 & 28 & 60 \\
\bottomrule
\end{tabular}
\caption{Agreement between WebVoyager (WV) and the independent
human annotator on the balanced 60-task audit sample.}
\label{tab:webvoyager-human-audit}
\end{table}

Human judgment agrees with WebVoyager on 28 of 30 automatically
successful cases (93.3\%; Wilson 95\% CI: 78.7--98.2\%) and
26 of 30 automatically unsuccessful cases (86.7\%; 95\% CI:
70.3--94.7\%). Overall agreement is 54/60, or 90.0\%
(95\% CI: 79.9--95.3\%), with Cohen's $\kappa=0.80$.
Here, $\kappa$ measures agreement between the human judgment and
WebVoyager rather than agreement between multiple human annotators.

The six disagreements consist of two cases judged successful by
WebVoyager but unsuccessful by manual review, and four cases judged
unsuccessful by WebVoyager but successful by manual review.
Because the audit intentionally contains equal numbers of
WebVoyager successes and failures, the 90.0\% value is a balanced
agreement rate. Weighting the two strata according to the full
\tool{} outcome distribution gives an estimated agreement of
91.7\%.

\subsection{Audit of TeleGen-Only Successes}
\label{sec:telegen-only-audit}

We separately inspect 30 tasks sampled from the 106 cases where
\tool{} succeeds but No-Telemetry repair fails. This audit examines
whether the measured gains correspond to functional repairs or
primarily to changes that make an already-correct outcome easier
for the browser agent or evaluator to observe.

We categorize 24 cases as genuine functional repairs, two as mixed
functional and judgeability improvements, and four as
judgeability-only improvements.

\begin{table}[t]
\centering
\footnotesize
\setlength{\tabcolsep}{5pt}
\renewcommand{\arraystretch}{1.08}
\begin{tabular}{lr}
\toprule
\textbf{Category} & \textbf{Count} \\
\midrule
Functional repair & 24 \\
Mixed functional/judgeability & 2 \\
Judgeability-only & 4 \\
\bottomrule
\end{tabular}
\caption{Manual categorization of 30 audited \tool{}-only successes.}
\label{tab:telegen-only-audit}
\end{table}

Functional repairs change application behavior required to complete
the evaluated task. Mixed cases combine a functional change with
clearer evaluator-facing feedback, while judgeability-only cases
make an already successful application state easier to operate or
observe. Thus, 24 of the 30 audited cases are clearly functional
repairs, and 26 involve at least some functional repair. The audit
also shows that judgeability contributes to a minority of the
measured gains.

\subsection{Cost and Runtime Analysis}
\label{sec:cost-analysis}

We additionally analyze the computational overhead of runtime
telemetry over all 101 WebGen-Bench projects. Table~\ref{tab:cost-runtime}
reports the recorded model calls, API cost, and project-level
wall-clock time for the three repair settings. This monetary
accounting is reported separately from the token statistics in
Table~3, which retain the aggregation used in the original
experiments.

Compared with No-Telemetry repair, \tool{} increases the average
number of model calls from 24.0 to 29.5 and the API cost from
\$0.28 to \$0.41 per project. Thus, the additional model-side cost
is approximately \$0.12 per project. Raw-Telemetry has a similar
cost of \$0.42 per project. The recorded repair-stage wall-clock
times are 711, 747, and 710 seconds per project for No-Telemetry,
Raw-Telemetry, and \tool{}, respectively.

\begin{table}[t]
\centering
\footnotesize
\setlength{\tabcolsep}{4.0pt}
\renewcommand{\arraystretch}{1.05}
\begin{tabular}{lrrr}
\toprule
\textbf{Setting} & \textbf{Calls} & \textbf{Cost (\$)} & \textbf{Time (s)} \\
\midrule
No-Tel.     & 24.0 & 0.28 & 711 \\
Raw-Tel.    & 29.9 & 0.42 & 747 \\
\tool{}     & 29.5 & 0.41 & 710 \\
\bottomrule
\end{tabular}
\caption{Mean per-project cost and runtime accounting on
WebGen-Bench. Cost is API cost under the DeepSeek-V4-Flash pricing
used in this analysis; time is recorded repair-stage wall-clock time.}
\label{tab:cost-runtime}
\end{table}

The observability pipeline also has modest instrumentation and
evidence overhead. Instrumentation processes 8.9 source files and
modifies 7.6 files per project on average, with no recorded
patch-application errors. Telemetry collection is integrated into
the instrumented WebVoyager execution: No-Telemetry uses one clean
WV1 execution for task feedback, whereas \tool{} uses one
instrumented WV1 execution that produces both task feedback and
runtime telemetry, so telemetry collection does not add another
full browser-execution pass. The resulting sanitized telemetry
report averages 52.9 KB per project and is compressed to a
477-byte telemetry brief, corresponding to a 99.1\% reduction in
evidence size.

\begin{table}[t]
\centering
\small
\begin{tabular}{llrrr}
\toprule
\textbf{Base} & \textbf{Level} & \textbf{Gain} & \textbf{Loss} & \textbf{Net} \\
\midrule
\multirow{2}{*}{No-tel.}
& Project & 48 & 16 & +32 \\
& Task    & 106 & 51 & +55 \\
\multirow{2}{*}{No brief}
& Project & 34 & 14 & +20 \\
& Task    & 65 & 56 & +9 \\
\bottomrule
\end{tabular}
\caption{Project- and task-level comparison between \tool{} and each
baseline. Gain denotes cases where telemetry-guided repair performs
better, Loss denotes cases where the baseline performs better, and
Net is Gain minus Loss.}
\label{tab:app-gain-loss}
\end{table}

The remaining projects or tasks have the same outcome under both settings. Against No-telemetry repair, telemetry-guided repair yields a net gain of 32 projects and 55 tasks. Against No brief, it yields a net gain of 20 projects and 9 tasks. These results show that the benefit of Observable repair is positive at both granularities: compared with both a no-runtime-telemetry baseline and a raw-log baseline, telemetry-guided repair improves more projects and tasks than it hurts.

\begin{table}[t]
\centering
\footnotesize
\setlength{\tabcolsep}{2.0pt}
\renewcommand{\arraystretch}{1.05}
\begin{tabular}{@{}lrrr@{}}
\toprule
\textbf{Pair (A/B)} & \textbf{A-only} & \textbf{B-only} & \textbf{$p$} \\
\midrule
No-Tel./\tool{}      & 51 & 106 & $1.35{\times}10^{-5}$ \\
Raw-Tel./\tool{}     & 56 & 65  & 0.4672 \\
Orig.P@2/\tool{}P@2  & 45 & 126 & $4.49{\times}10^{-10}$ \\
\bottomrule
\end{tabular}
\caption{Two-sided exact McNemar tests on paired task outcomes.
A-only and B-only denote tasks solved exclusively by the first and
second settings in each pair.}
\label{tab:paired-significance}
\end{table}

\section{Implementation Details}
\label{app:implementation-details}

All repair settings use a single repair round. This design keeps the comparison controlled: after the initial browser-based evaluation, each setting receives one opportunity to modify the generated application. We use the same task set and evaluation protocol across settings.

We separate the text/code model from the browser-execution model. The text/code model is responsible for initial generation, telemetry instrumentation, telemetry-brief construction, and repair. In the main experiments, this model is DeepSeek-V4-Flash. All WebVoyager model-driven steps, including browser exploration, task execution, and final task evaluation, are routed to Qwen3.5-Plus. Additional text/code model results are reported in Appendix~\ref{app:model-generalization}.

Telemetry instrumentation is performed as an LLM-based patching step. The instrumentation model is instructed to add runtime telemetry only and not to repair functional behavior during this stage. To reduce uncontrolled edits, the model outputs code patches rather than directly modifying the workspace. These patches are then applied locally before executing the instrumented application.

The no-telemetry baseline receives the task description, source code, and browser-level task feedback, but no runtime logs. The no-brief setting receives raw runtime logs collected from the instrumented application without telemetry sanitization and compression. Observable repair receives the compressed telemetry brief, which removes duplicated or irrelevant logs and preserves task-relevant observations such as navigation, user interactions, network requests, runtime errors, and state changes.

For observable repair, we first execute the instrumented application with WebVoyager and collect runtime observations during browser-based task execution. The collected logs are then sanitized and compressed into a repair-oriented telemetry brief. The repair agent receives this brief together with the task feedback and source code, and produces a repair patch. The repaired application is evaluated using the same WebVoyager protocol as the other settings.

For RQ2, token usage is aggregated at the project level for each repair setting. We separately record input and output tokens and report average per-project usage in the main text. The no-brief and observable settings include the additional runtime evidence provided to the repair agent, while the no-telemetry setting does not.

\paragraph{Computational budget.}
All experiments use proprietary API-based models. DeepSeek-V4-Flash is used as the main text/code model, while Qwen3.5-Plus is used for WebVoyager execution and screenshot-based task judgment. The exact parameter counts of these proprietary models are not publicly disclosed. We do not train or fine-tune any model. The main computational cost comes from API calls during generation, telemetry instrumentation, telemetry-brief construction, repair, and browser-based evaluation. We report token usage for the main repair settings in Table~\ref{tab:rq2-token-cost}. Web application execution and browser testing are run in a local sandboxed environment using the benchmark execution protocols.

\section{Prompt Templates}
\label{app:prompts}

This section summarizes the prompt templates used in our pipeline.
The actual prompts include project-specific instructions, file paths, source code, raw telemetry logs, task feedback, and task metadata.
For readability, we replace such instance-specific content with placeholders while preserving the key instructions and constraints.

\subsection{Initial Generation Prompt}
\label{app:prompt-initial-generation}

The initial generation prompt asks the text/code model to generate a complete runnable web application from the website specification and UI task suite.
This step follows the original generation protocol of the benchmark.
It does not include telemetry instrumentation; telemetry is inserted later on a separate copy of the generated v1 application.

\begin{promptblock}
You are a professional full-stack developer.
Generate a complete, working web application from the provided website specification and UI task suite.

## Project ID
{{PROJECT_ID}}

## Website Specification
{{WEBSITE_SPECIFICATION}}

## Category Information
- Primary Category: {{PRIMARY_CATEGORY}}
- Subcategories: {{SUBCATEGORIES}}

## UI Task Suite
The generated application will be evaluated with the following UI tasks and expected results.

{{UI_TASK_SUITE}}

## Technical Requirements
Generate a full-stack application with:
1. Backend: Node.js/Express RESTful API on port 5001
2. Frontend: React + Vite on port 3000
3. CORS enabled between frontend and backend
4. All required dependencies declared in package.json files
5. Setup scripts and clear run instructions

## Implementation Requirements
1. Create the project directory structure.
2. Implement the backend routes, controllers, and data logic needed by the website specification.
3. Implement the frontend pages, components, navigation, state updates, forms, and user interactions needed by the UI tasks.
4. Ensure frontend API calls match backend routes, methods, and payload formats.
5. Prepare the environment and perform basic syntax or smoke checks when possible.
6. Generate complete working code with no placeholders or TODOs.

## Telemetry Constraint
Do not add telemetry instrumentation, debug logging, or semantic tracing markers during initial generation.
Runtime telemetry will be inserted later by a separate instrumentation step.

Generate COMPLETE, WORKING code.
\end{promptblock}

\subsection{Telemetry Instrumentation Prompt}
\label{app:prompt-instrumentation}

Telemetry instrumentation is performed as an LLM-based patching step on a separate copy of the v1 application.
The instrumentation model receives the task context and one source file at a time, and returns structured patch operations rather than directly modifying the workspace.
This design allows us to inspect and apply instrumentation locally, while preserving the clean v1 source code for repair.

\begin{promptblock}
You are a code instrumentation assistant for {{LANGUAGE}}.

Your job is to add lightweight runtime telemetry logs that help diagnose the tested web application flows.
You must not rewrite the full file.
You must return PATCH OPERATIONS ONLY as JSON.

## Goal
Add telemetry that makes task-relevant runtime behavior visible during browser-based execution.
Use the website specification, UI task suite, and expected behaviors to infer important user journeys and likely code paths.

Add logs only at meaningful points, such as:
- navigation entry points
- click and input event handlers
- form submissions
- state updates
- fetch or axios requests
- network response handling
- route handlers
- controller entry points
- critical branch decisions
- error handling paths

Prefer consistent telemetry prefixes:
- [Telemetry] Interaction
- [Telemetry] Network Request
- [Telemetry] Network Response
- [Telemetry] StateChange
- [Telemetry] Branch
- [Telemetry] Error

## Strict Instrumentation Rules
1. Instrumentation is for observation only.
2. Do not change functional behavior.
3. Do not repair bugs.
4. Do not refactor unrelated logic.
5. Do not remove existing code.
6. Do not invent APIs, routes, variables, or data structures not supported by the current file.
7. Keep patches minimal.
8. Prefer adding logging lines or the smallest structural change required to place logging safely.
9. Do not modify imports unless required for a safe logging pattern.
10. If this file is not relevant to the tested flows, return {"operations": []}.
11. At most 8 operations.

## Output Format
Return exactly one JSON object.
Do not include markdown or prose.

JSON schema:
{
  "operations": [
    {
      "action": "replace",
      "search": "exact existing snippet from the file",
      "replace": "replacement snippet with added telemetry logs",
      "reason": "short reason"
    }
  ]
}

Every search snippet must appear exactly as-is in the current file and match exactly once.
Include enough surrounding code in the search snippet to make repeated lines unique.

## Task Context
{{TASK_CONTEXT}}

## Current File
Path: {{FILE_PATH}}

## Current File Content
{{FILE_CONTENT}}
\end{promptblock}

The task context supplied to the instrumentation model contains the website specification, project category metadata, and the UI task suite with expected behaviors.
This context is used only to identify task-relevant execution paths for telemetry, not to repair the application.

\begin{promptblock}
Project ID: {{PROJECT_ID}}

Website specification:
{{WEBSITE_SPECIFICATION}}

Primary category: {{PRIMARY_CATEGORY}}
Subcategories: {{SUBCATEGORIES}}

UI tasks and expected behaviors:
1. Task: {{TASK_1}}
   Expected result: {{EXPECTED_RESULT_1}}

2. Task: {{TASK_2}}
   Expected result: {{EXPECTED_RESULT_2}}

...

Use these tasks to infer the likely user journey, navigation path, frontend interactions, backend endpoints, network behavior, and state transitions that should be logged.
\end{promptblock}

\subsection{Telemetry Brief Construction Prompt}
\label{app:prompt-brief}

The telemetry brief construction prompt compresses raw telemetry logs into a concise, repair-oriented summary.
It removes duplicated or irrelevant entries while preserving task-relevant runtime evidence.
The brief is not intended to be a complete execution transcript; it summarizes what happened during execution in a form that is useful for repair.

\begin{promptblock}
You are a telemetry analysis assistant.

You will receive:
1. Task context
2. Task feedback
3. Raw telemetry logs collected during browser-based execution

Your job is to clean noisy logs, extract meaningful runtime evidence, and reconstruct a concise execution summary for the repair agent.

Do NOT provide code patches.
Do NOT provide implementation instructions.
Do NOT speculate beyond the evidence.
Only describe what happened factually and objectively.

## Discard or Compress
- Framework noise
- Browser extension messages
- Development server messages unrelated to the task
- Exact duplicate log entries
- Repeated low-level messages that do not change the execution state
- Third-party library noise
- Logs unrelated to the tested task flow

## Retain and Organize
Preserve task-relevant evidence, including:
- user actions and interaction events
- navigation attempts
- network requests and responses
- JavaScript exceptions
- backend errors
- state changes
- missing or stalled interactions
- data loaded but not rendered
- successful requests followed by no visible update
- source file and line references when available

## Output Format

### Task Context
Briefly summarize the task being executed and the expected result.

### Task Feedback
Summarize the observed pass/fail outcome and any task executor feedback.

### User Operation Timeline
Provide a chronological timeline of task-relevant events.

Use event types such as:
- USER_ACTION
- NAVIGATION
- API_REQUEST
- API_RESPONSE
- STATE_CHANGE
- ERROR
- MISSING_INTERACTION
- OBSERVED_GAP

### Runtime Evidence Summary
Summarize the most important runtime evidence.
Focus on evidence that helps distinguish why the task failed.

### Affected Code Locations
List source files, functions, routes, components, or line numbers referenced in the logs, if available.

### Compression Report
Report the approximate number of raw entries, retained entries, and discarded or compressed entries.

## Constraints
- Use only information present in the task context, task feedback, and raw telemetry logs.
- Do not invent missing code locations.
- Do not recommend specific patches.
- Do not include sections titled Root Cause or Recommendations.
- Preserve precise file references when present.

## Task Context
{{TASK_CONTEXT}}

## Task Feedback
{{TASK_FEEDBACK}}

## Raw Telemetry Logs
{{RAW_TELEMETRY_LOGS}}
\end{promptblock}

\subsection{Repair Prompt Variants}
\label{app:prompt-repair-variants}

The three repair prompts differ only in the runtime evidence available to the repair agent.
No-telemetry repair receives task feedback without runtime telemetry.
No-brief repair receives raw telemetry logs.
\tool{} receives the compressed telemetry brief.
All repair variants edit the clean v1 source code rather than the instrumented copy.

\subsubsection{No-Telemetry Repair Prompt}

\begin{promptblock}
# Core Objective
You are an expert full-stack implementation agent.
The mounted workspace contains the clean v1 source code of a generated decoupled frontend/backend web application.

Your goal is to maximize the number of task executor tasks that pass using only:
1. the website specification,
2. the task descriptions and expected results,
3. the v1 task feedback,
4. static source-code inspection.

# Baseline Constraint
This is the no-telemetry repair branch.
Do not rely on telemetry_brief.md or telemetry_report.md.
Do not add telemetry instrumentation, debug logging, or tracing as part of the repair.

# Analysis Strategy
1. Read the task evidence section. It lists the exact tasks, expected results, and v1 pass/fail observations.
2. Inspect the frontend and backend code to identify missing features, broken navigation, bad API contracts, state-update issues, or runtime-prone logic.
3. Compare the expected behavior with the current implementation.
4. Implement the minimum targeted changes needed for task completion.
5. Preserve the existing application structure and technology choices.

# Execution Constraints
1. Improve task completion, not just superficial errors.
2. Do not add logging or instrumentation.
3. Do not refactor technology stacks or file structures.
4. Ensure frontend API calls match backend routes, methods, and payloads.
5. Do not run sandbox tests.
6. Finish once all targeted modifications are complete.
\end{promptblock}

\subsubsection{No-Brief Repair Prompt}

\begin{promptblock}
# Core Objective
You are an expert full-stack implementation agent.
The mounted workspace contains the clean v1 source code of a generated decoupled frontend/backend web application.

Your goal is to maximize the number of task executor tasks that pass.

In addition to the standard task feedback, you are given raw telemetry logs collected from a separately instrumented copy of the v1 application.
These logs provide runtime evidence about what happened during browser-based execution.
They may be noisy, duplicated, or incomplete, so they should be interpreted together with the task feedback and source code.

# Evidence Available
You may use:
1. website specification,
2. task descriptions and expected results,
3. v1 task feedback,
4. clean v1 source code,
5. telemetry_report.md containing raw telemetry logs.

# Clean Source Constraint
The current workspace is the clean v1 source code used for repair.
Do not assume that telemetry instrumentation is part of the code being repaired.
Do not add new telemetry instrumentation as part of the repair unless it is strictly necessary for application functionality, which should normally not be the case.

# Analysis Strategy
Step 1: Understand what should work.
Read the task evidence section, including task descriptions, expected results, and pass/fail observations.

Step 2: Understand what actually happened.
Read telemetry_report.md.
Use raw telemetry logs to identify:
- HTTP errors
- JavaScript exceptions
- failed or missing interactions
- navigation attempts
- network requests and responses
- state changes
- frontend-backend data mismatches
- events that occurred internally but did not produce the expected UI result

Step 3: Gap analysis.
Compare expected behavior with observed runtime behavior and source-code implementation.

Step 4: Implement fixes.
Prioritize:
1. missing task features,
2. runtime errors,
3. broken navigation paths, routes, links, and buttons,
4. frontend-backend request/response mismatches,
5. state updates that fail to render expected UI changes.

# Execution Constraints
- Make targeted changes only.
- Do not add telemetry instrumentation as a repair strategy.
- Do not refactor technology stacks or file structures.
- Do not run sandbox tests.
- Finish once all targeted modifications are complete.
\end{promptblock}

\subsubsection{\tool{} Repair Prompt}

\begin{promptblock}
# Core Objective
You are an expert full-stack implementation agent.
The mounted workspace contains the clean v1 source code of a generated decoupled frontend/backend web application.

Your goal is to maximize the number of task executor tasks that pass.

In addition to the standard task feedback, you are given a telemetry brief produced from raw telemetry logs collected on a separately instrumented copy of the v1 application.
The telemetry brief summarizes task-relevant runtime evidence and should help you localize failures before editing code.

# Evidence Available
You may use:
1. website specification,
2. task descriptions and expected results,
3. v1 task feedback,
4. clean v1 source code,
5. telemetry_brief.md.

# Clean Source Constraint
The current workspace is the clean v1 source code used for repair.
Do not edit the instrumented copy.
Do not add telemetry instrumentation, debug logging, or tracing as part of the repair.

# Analysis Strategy
Step 1: Understand what should work.
Read the task evidence section, including the exact tasks, expected results, and v1 pass/fail observations.

Step 2: Understand what actually happened.
Read telemetry_brief.md.
Use the telemetry brief to identify:
- failed or missing interactions
- navigation attempts that did not reach the expected page
- successful requests followed by missing UI updates
- JavaScript exceptions
- backend errors
- frontend-backend data mismatches
- state changes that did not produce the expected visible result
- task-relevant code locations mentioned in the brief

Step 3: Gap analysis.
Compare the expected behavior, v1 task feedback, telemetry evidence, and source code.
Identify missing features, broken UI flows, failed actions, data contract mismatches, and runtime errors.

Step 4: Implement fixes.
Prioritize:
1. missing task features,
2. runtime errors,
3. broken routes, links, buttons, and entry points,
4. frontend-backend request/response mismatches,
5. state updates or rendering logic needed to complete the task.

# Execution Constraints
- Make minimal targeted changes.
- Do not add telemetry instrumentation.
- Do not refactor technology stacks or file structures.
- Do not run sandbox tests.
- Finish once all targeted modifications are complete.
\end{promptblock}

\subsection{Task Evidence Block}
\label{app:prompt-e2e-evidence}

All repair prompts receive a dynamically constructed task evidence block describing the application purpose, expected task behavior, and v1 task executor outcomes.

\begin{promptblock}
---
# Task Evidence from v1 Baseline

The following information describes what the application is supposed to do and what actually happened during automated task execution of the v1 application.

## Application Purpose
{{WEBSITE_SPECIFICATION}}

## Expected Behaviors
### Expected Task 1
Task: {{TASK_1}}
Expected Result: {{EXPECTED_RESULT_1}}

### Expected Task 2
Task: {{TASK_2}}
Expected Result: {{EXPECTED_RESULT_2}}

...

## v1 Baseline Task Results
Summary:
{{N_PASSED}} passed, {{N_FAILED}} failed, {{N_UNKNOWN}} unknown or timeout

### {{TASK_NAME}} -- {{VERDICT}}
Task executor observation: {{OBSERVATION}}

...
\end{promptblock}

\subsection{WebVoyager Execution and Evaluation Prompts}
\label{app:prompt-webvoyager}

For WebGen-Bench, WebVoyager is used both as a browser-interaction task executor and as a screenshot-based evaluator.
The following templates summarize the browsing and evaluation prompts.

\subsubsection{Browsing Agent Prompt}

\begin{promptblock}
Imagine you are a robot browsing the web, just like humans.
In each iteration, you receive an observation with a screenshot of a webpage and text context.
The screenshot contains numerical labels placed at web elements.

Analyze the screenshot and choose exactly one action:
- Click [Numerical_Label]
- Type [Numerical_Label]; [Content]
- Scroll [Numerical_Label or WINDOW]; [up or down]
- Wait
- GoBack
- ANSWER; [content]

Key rules:
1. Execute only one action per iteration.
2. Do not repeat the same ineffective action if the webpage remains unchanged.
3. Use ANSWER only after completing all required steps.
4. Pay attention to search boxes, buttons, filters, sort controls, dates, forms, and labeled web elements.
5. Focus on the numerical labels attached to web elements.
6. When a task requires checking a final state, navigate and interact until the expected result is visible or clearly unavailable.

Your reply must follow:
Thought: {{BRIEF_REASONING}}
Action: {{ONE_ACTION}}
\end{promptblock}

\subsubsection{Evaluation Prompt}

\begin{promptblock}
As an evaluator, you will be presented with:
1. Web task instruction
2. Result screenshots
3. Result response

Your responsibility is to assess whether the result satisfies the task instruction.
You do not need to interact with web pages.
Do not assume information not present in the screenshot or response.
If the screenshot and response conflict, the screenshot prevails.

Evaluate whether the task has been successfully accomplished.
Explain the evidence for your evaluation, then provide a final verdict:
SUCCESS or NOT SUCCESS.

Input:
TASK: {{TASK_INSTRUCTION}}
Result Response: {{AGENT_ANSWER}}
Screenshots: {{SCREENSHOTS}}
\end{promptblock}

\end{document}